\documentclass[final]{elsarticle}
\usepackage{natbib}
\usepackage{graphicx}
\usepackage{textcomp}
\usepackage{lineno,hyperref}
\usepackage{amsmath}

\journal{Journal of Computational and Applied Mathematics, Elsevier}

\begin{document}
	
\begin{frontmatter}

\title{The link to the formal publication is via\\ {\small \url{https://doi.org/10.1016/j.cam.2006.12.031}}\\ .\\ Multigrid Solvers in Reconfigurable Hardware}

\author[SafaaAddress]{Safaa Kasbah\corref{mycorrespondingauthor}}

\author[IssamAddress]{Issam Damaj}
\author[RamziAddress]{Ramzi Haraty}
\cortext[mycorrespondingauthor]{Corresponding Author}

\address[SafaaAddress]{Computer Science and Mathematics Department, Lebanese American University, Beirut, Lebanon, safaakasbah@gmail.com}

\address[IssamAddress]{Electrical and Computer Engineering Department, Dhofar University, Salalah, Sultanate of Oman, damaj@du.edu.om}

\address[RamziAddress]{Computer Science and Mathematics Department, Lebanese American University, Beirut, Lebanon, rharaty@lau.edu.lb}

%\author[LAU]{Safaa J. Kasbah\thanksref{sjk}}
%\author[Dohfar]{Issam W. Damaj \thanksref{iwd}}
%\author[LAU]{Ramzi A. Haraty \thanksref{rah}}

%\address[LAU]{Division of Computer Science and Mathematics, Lebanese American University, Beirut, Lebanon}
%
%\address[Dohfar]{Department of Electrical and Computer Engineering, Dhofar University, Salalah, Sultanate of Oman}
%\thanks[sjk]{safaa.kasbah@lau.edu.lb}
%\thanks[iwd]{$i\textunderscore damaj@du.edu.om$}
%\thanks[rah]{rharaty@lau.edu.lb}

\begin{abstract}
The problem of finding the solution of Partial Differential
Equations (\textit{PDEs}) plays a central role in modeling real
world problems. Over the past years, Multigrid solvers have showed
their robustness over other techniques, due to its high convergence
rate which is independent of the problem size. For this reason, many
attempts for exploiting the inherent parallelism of Multigrid have
been made to achieve the desired efficiency and scalability of the
method. Yet, most efforts fail in this respect due to many factors
(time, resources) governed by software implementations. In this
paper, we present a hardware implementation of the V-cycle Multigrid
method for finding the solution of a 2D-Poisson equation. We use
\textit{Handel-C} to implement our hardware design, which we map
onto available Field Programmable Gate Arrays (\textit{FPGAs}). We
analyze the implementation performance using the \textit{FPGA}
vendor's tools. We demonstrate the robustness of Multigrid over
other iterative solvers, such as Jacobi and Successive Over
Relaxation (\textit{SOR}), in both hardware and software. We compare
our findings with a \textit{C++} version of each algorithm. The
obtained results show better performance when compared to existing
software versions.
\end{abstract}

\begin{keyword}
2-D Poisson; \textit{FPGA}; Reconfigurable Computing
\end{keyword}

\end{frontmatter}

\section{Introduction}
Physical, chemical and biological phenomena are modeled using
Partial Differential Equations (\textit{PDEs}). Interpreting and
solving (\textit{PDEs}) is the key for understanding the behavior of
the modeled system. The broad field of modeling real systems has
drawn the researchers' attention for designing efficient algorithms
for solving (\textit{PDEs}). The Multigrid method has been shown to
be the fastest method due to its high convergence rate which is
independent from the problem size. However, the computation of such
solvers is complex and time consuming. Many attempts for exploiting
the inherent parallelism of Multigrid have been made to achieve the
desired efficiency and scalability of the method. Yet, most efforts
fail in this respect due to many factors (time and resources)
governed by software implementations upon parallelizing the
algorithm.

Over the past years, researchers have benefited from the continuous
advances in hardware devices and software tools to accelerate the
computation of complex problems [2]. At early stages, algorithms
were designed and implemented to run on a general purpose processor
(software). Techniques for optimizing and parallelizing the
algorithm, when possible, were then devised to achiever better
performance. As applications get more complex, the performance
provided by processors degenerates. A better performance could be
achieved using a dedicated hardware where the algorithm is digitally
mapped onto a silicon chip, Integrated Circuit (\textit{IC}). Though
it provides better performance than the processor technology, the
\textit{IC} technology (hardware) lacks flexibility.

In the last decade, a new computing paradigm, Reconfigurable
Computing (\textit{RC}), has emerged [14]. \textit{RC}-systems
overcome the limitations of the processor and the \textit{IC}
technology. \textit{RC}-systems benefit from the flexibility offered
by software and the performance offered by hardware [23], [14].
\textit{RC} has successfully accelerated a wide variety of
applications including cryptography and signal processing [22]. This
achievement requires a reconfigurable hardware, such as an
\textit{FPGA}, and a software design environment that aids in the
creation of configurations for the reconfigurable hardware [14].

In this paper, we present a hardware implementation of the V-cycle
Multigrid algorithm for the solution of a 2D-Poisson equation using
different classes of \textit{FPGAs}: \textit{Xilinx Virtex II Pro},
\textit{Altera Stratix} and \textit{Spartan3L} which is embedded on
the RC10 board from \textit{Celoxica}. We use \textit{Handel-C}, a
higher-level hardware design language, to code our design which is
analyzed, synthesized, and placed and routed using the
\textit{FPGAs} proprietary software (\textit{DK} Design Suite,
\textit{Xilinx ISE 8.1i and Quartus II 5.1}). We demonstrated the
robustness of the Multigrid algorithm over the Jacobi and the
\textit{SOR} algorithms, in both hardware and software. We compare
our implementations results with existing software version of each
algorithm, since there are no hardware implementations of
\textit{MG, Jacobi} and \textit{SOR} in the literature.

The rest of the paper is organized as follows: In Sections 2 and 3,
we present a general overview of Multigrid solvers and
Reconfigurable Computing, respectively. In Section 4, we describe
our hardware implementation of the V-cycle \textit{MG} for the
solution of 2D-Poisson equation. Then, the implementation results
are presented in Section 5, where we: a) report \textit{MG} results,
b) compare these results with a software version written in C++ and
running on a general purpose processor, c) report \textit{Jacobi}
and \textit{SOR} hardware implementation results and compare them
with their software versions, d) compare the results obtained in a)
and c) showing how \textit{MG} outperforms \textit{Jacobi} and
\textit{SOR}, in both hardware and software versions. Section 6
concludes the work and addresses possible directions to future work.

\section{Multigrid Solvers}

Multigrid methods are fast linear iterative solvers used for finding
the optimal solution of a particular class of partial differential
equations. Similar to classical iterative methods (\textit{Jacobi},
Successive Over Relaxation (\textit{SOR}) and Gauss Seidel), an
\textit{MG} method "\textit{starts with an approximate solution to
the differential equation; and in each iteration, the difference
between the approximate solution and the exact solution is made
smaller}" [7].

In general, the error resulting from the exact and approximate
solution will have components of different wavelengths:
high-frequency components and low-frequency components [7].
Classical iterative methods reduce high-frequency/ oscillatory
components of error rapidly, but reduce low-frequency/smooth
components of error much more slowly [42].

The Multigrid strategy overcomes the weakness of classical iterative
solvers by observing that components that appear smooth on fine grid
may appear oscillatory when sampled on coarser grid [8]. The
high-frequency components of the error are reduced by applying any
of the classical iterative methods. The low-frequency components of
error are reduced by a coarse-grid correction procedure [11], [42].

A Multigrid cycle starts by applying any classical iterative method
(Jacobi, Gauss Seidel or Successive Over Relaxation) to find an
approximate solution for the system. The Residual operator is then
applied to find the difference between the actual solution and the
approximate solution. The result of this operator measures the
goodness of the approximation. Since it is easier to solve a problem
with less number of unknowns [9], [30], a special
operator-Restriction- for mapping the residual to a coarser grid
(less number of unknowns) - is applied for several iterations until
the scheme reaches the bottom of the grid hierarchy. Then, the
coarse grid solver operator is applied to find the error on the
coarsest grid. Afterwards, the interpolation operator is applied to
map the coarse grid correction to the next finer grid in an attempt
to improve the approximate solution. This procedure is applied until
the top grid level is reached giving a solution with residual zero.
Finishing with several iterations back to the finest grid gives a
so-called- V-cycle Multigrid [11], [18], [25].

\begin{figure}[h]
\begin{center}
\includegraphics{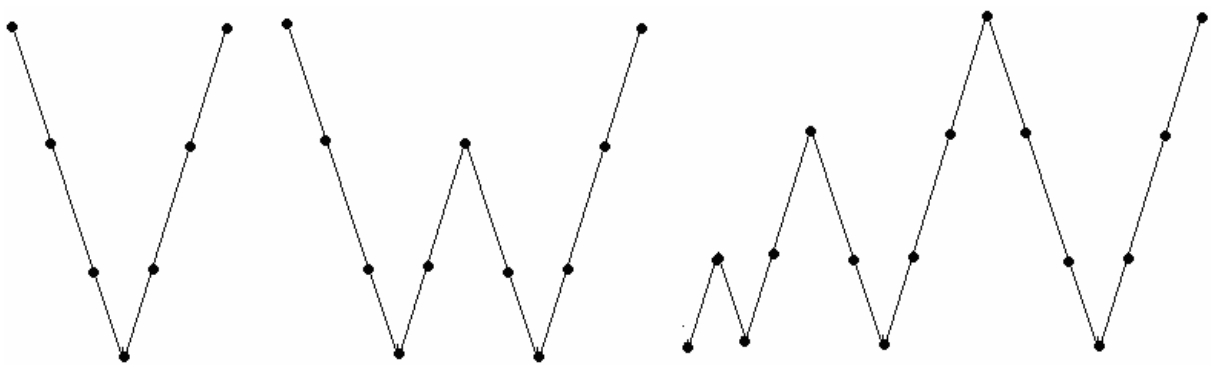}
\caption{V-cycle, W-cycle, and Full-cycle \textit{MG}  \label{f1}}
\end{center}
\end{figure}

\subsection {Multigrid Components}
A Multigrid algorithm uses five algorithmic components:
Smoother/Relaxation, Residual computation, Restriction, Coarse grid
solver, Interpolation.

\textbf{Relaxation/Smoother}: Is the most important component of
Multigrid algorithm. This component is responsible for generating an
approximate solution by reducing-smoothing/relaxing- the high
frequency error component of the solution imposed upon approximating
the solution.

Iterative methods that can be applied as the smoother in the
Multigrid solver include: Gauss-Seidel, \textit{Jacobi}, and
Successive Over Relaxation (\textit{SOR}) [7], [19] .The
Gauss-Seidel method has the fastest convergence rate and is thus the
best candidate. The Gauss-Seidel method can be used in both the
pre-smoothing and post-smoothing steps. When used as a pre-smoother,
the Gauss-Seidel method is responsible for reducing the
high-frequency error components and getting a smoother error [8].
When used as a post-smoother, the Gauss-Seidel method is responsible
for removing new high-frequency error component that might be
produced by the coarse grid correction and interpolation.

In general, Gauss-Seidel method can generate a solution using
\ref{eq1}
\begin{equation}
\label{eq1} x^{k}_{i}=\frac{b_{i}}- \sum_{j<i}a_{i,j} x^{(k)}_{j}
-\sum_{j>i}a_{i,j}{} x^{(k-1)}_{j}{a_{i,j}}
\end{equation}

When used as a post-smoothing and pre-smoothing steps in the
Multigrid method, the solution is in the form:
\begin{equation}
\label{eq2}
u^{t+1}_{i,j}=\frac{1}{4}(t_{i+1,j}+t_{i-1,j}+t_{i,j-1}+t_{i,j+1}+h^{2}f_{i,j})
\end{equation}

where $i$ and $j$ are the row and column indices of the gird (Barret
et. al., 1994).

\textbf{Residual Computation}: Let $\hat{u}$ be an approximate
solution to the exact solution $u$, the residual is defines as:

\begin{equation}
\label{eq3} r_{h}=f_{h}-A_{h}u_{h}^{(v_{1})}=A_{h}e_{h}
\end{equation}
where  $e=u-\hat{u}$ is the error.

The residual must be computed before it can be restricted to the
coarser grid.

\textbf{Restriction}: This component is responsible for transporting
the residual of the fine grid  to the coarser grid . There are
various techniques for restricting the residual; full weighing, half
weighting and injection [30].

Let $H=2h$ be the mesh size on a finer grid.\\ Let $r_{H}$  be the
prolongation of coarser-grid residual to the finer grid. \\Let
$I^{H}_{h}$ be a prolongation, from $h\rightarrow H$ (bilinear in
our case),
\\Now, We can compute $r_{H}$ using: $r_{H}=I^{H}_{h}r_{h}$

\textbf{Course grid solver}: This operator, often called coarse-grid
correction is performed on the coarse grid. Applying this operator
along with the smoothing operator has a substantial effect on the
reduction of the residual for all frequencies.  However, the coarse
grid solver is applied only on the coarsest grid making the cost of
this operator negligible to the overall computational cost of the
Multigrid method [44].

\begin{equation}
\label{eq5} u^{(v_{1}+1)}_{h}=u^{v_{1}}+ I_{H}^{h}e_{H}
\end{equation}

\textbf{Interpolation/Prolongation}: Transports the correction
obtained on the coarser grid to the fine grid. There are various
techniques to do so: e.g., bilinear and 7 point interpolation.

Let $h$ be the mesh size on a finer grid.\\ Let $r_{h}$  be the
prolongation of coarser-grid residual to the finer grid. \\Let
$I^{h}_{H}$ be a prolongation, from $H\rightarrow h$ (bilinear in
our case),
\\Now, We can compute $r_{h}$ using: $r_{h}=I^{h}_{H}r_{h}$

The simplest Multigrid algorithm is based on a two-grid improvement
scheme: fine grid and coarse grid. The fine grid, $\Omega^h$, with
$N=2^{l}+2$ points and the coarse grid, $\Omega^2h$ , with
$N=2^{l-1}+2$points.

In this work, we implement the V-cycle Multigrid to find the
solution of a 2-D Poisson equation. Briefly, the V-cycle Multigrid
algorithm starts with an initial approximation to the expected
solution, goes down to the coarsest grid, and then goes back to the
finest grid [11].

\begin{figure}[h]
\begin{center}
\includegraphics{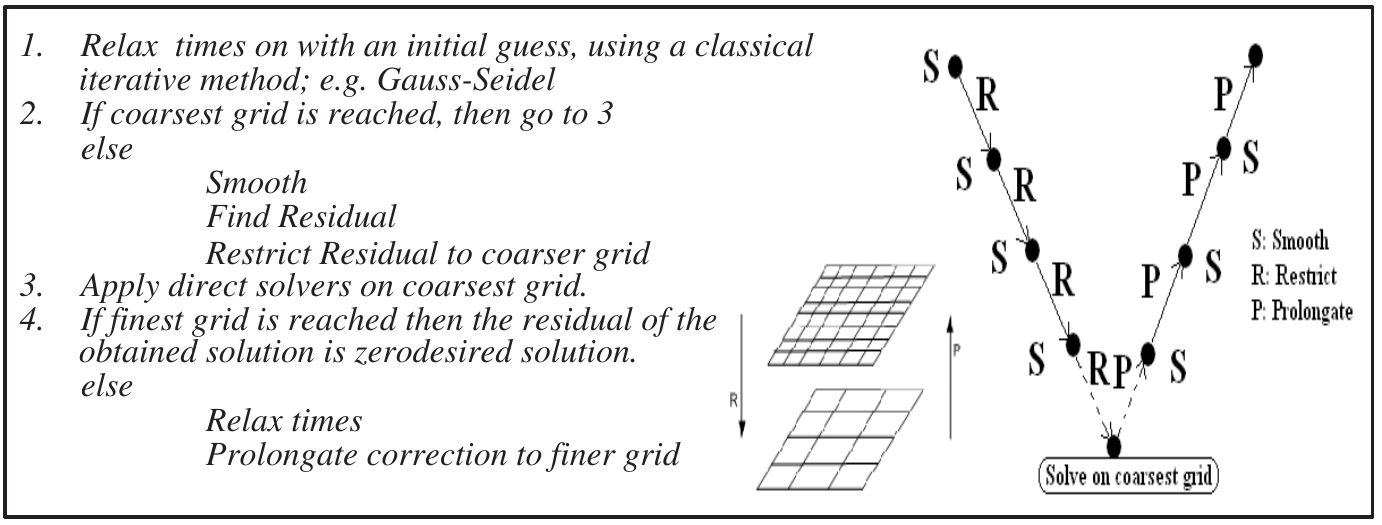}
\caption{V-cycle \textit{MG}\label{f2}}
\end{center}
\end{figure}

\subsection {Multigrid Solution of Poisson's Equation in 2-D}

The V-cycle Multigrid algorithm is applied to find the solution to a
2-D Poisson equation in the form:
\begin{equation}
\frac{\partial ^2 u(x,y)}{\partial x ^2}+ \frac{\partial ^2
u(x,y)}{\partial y ^2}= f_{x,y}
\end{equation}
or in the form  $\nabla^{2}u=f$ when written in vector notation
[35].

\section{Reconfigurable Computing}

Today, it becomes possible to benefit from the advantages of both
software and hardware with the presence of the Reconfigurable
Computing paradigm [14]. Actually, the first idea to fill the gap
between the two computing approaches, software and hardware,  goes
back to the 1960s when Gerald Estrin proposed the concept of
\textit{RC} [40].

The basic idea of Reconfigurable Computing is the "\textit{ability
to perform certain computations in hardware to increase the
performance, while retaining much of the flexibility of a software
solution}" [14].

Reconfigurable computing systems can be either of fine-grained or of
coarse-grained architecture. An \textit{FPGA} is a fine-grained
reconfigurable unit while a reconfigurable array processor is a
coarse-grained reconfigurable unit .In the fine-grained architecture
each bit can be configured; while in the coarse-grained
architecture, the operations and the interconnection of each
processor can be configured. Example of a coarse-grained system is
the \textit{MorphoSys} which is intended for accelerating data path
applications by combining a general purpose micro-processor and an
array of coarse grained reconfigurable cells [3].

The realization of the \textit{RC} paradigm is made possible by the
presence of programmable hardware such as large scale Complex
Programmable Logic Devices (\textit{CPLDs}) and Field Programmable
Gate Arrays (\textit{FPGAs}) [37]. Reconfigurable computing involves
the modification of the logic within the programmable device to
suite the application at hand.

\subsection{Hardware Compilation}

There are certain procedures to be followed before implementing a
design on an \textit{FPGA}. First, the user should prepare his/her
design by using either a schema editor or by using one of the
Hardware Description Languages \textit{(HDLs)} such as \textit{VHDL}
(Very high scale integrated circuit Hardware Description Language)
and \textit{Verilog}. With schema editors, the designer draws
his/her design by choosing from the variety of available components
(multiplexers, adders, resistors, ..) and connect them by drawing
wires between them. A number of companies supply schema editors
where the designer can drag and drop symbols into a design, and
clearly annotate each component [39]. Schematic design is shown to
be simple and easy for relatively small designs. However, the
emergence of big and complex designs has substantially decreased the
popularity of schematic design while increasing the popularity of
\textit{HDL} design. Using an \textit{HDL}, the designer has the
choice of designing either the structure or the behavior of his/ her
design. Both \textit{VHDL} and \textit{Verilog} support structural
and behavioral descriptions of the design at different levels of
abstractions. In structural design, a detailed description of the
system's components, sub-components and their interconnects are
specified. The system will appear as a collection of gates and
interconnects [39]. Though it has a great advantage of  having an
optimized design, structural presentation becomes hard, as the
complexity of the system increases. In behavioral design, the system
is considered as a black box with inputs and outputs only, without
paying attention to its internal structure [24]. In other words, the
system is described in terms of how it behaves rather than in terms
of its components and the interconnection between them. Though it
requires more effort, structural representation is more advantageous
than the behavioral representation in the sense that the designer
can specify the information at the gate-level allowing optimal use
of the chip area [41]. It is possible to have more than one
structural representation for the same behavioral program.

Noting that modern chips are too complex to be designed using the
schematic approach, we will choose the \textit{HDL} instead of the
schematic approach to describe our designs.

Whether the designer uses a schematic editor or an \textit{HDL}, the
design is fed to an Electronic Design Automation (\textit{EDA}) tool
to be translated to a netlist. The netlist can then be fitted on the
\textit{FPGA} using a process called place and route, usually
completed by the \textit{FPGA} vendors' tools. Then the user has to
validate the place and route results by timing analysis, simulation
and other verification methodologies. Once the validation process is
complete, the binary file generated is used to (re)configure the
\textit{FPGA} device. More about this process is found in the coming
sections.

Implementing a logic design on an \textit{FPGA} is depicted in the
figure below:
\begin{figure}[h]
\begin{center}
\includegraphics{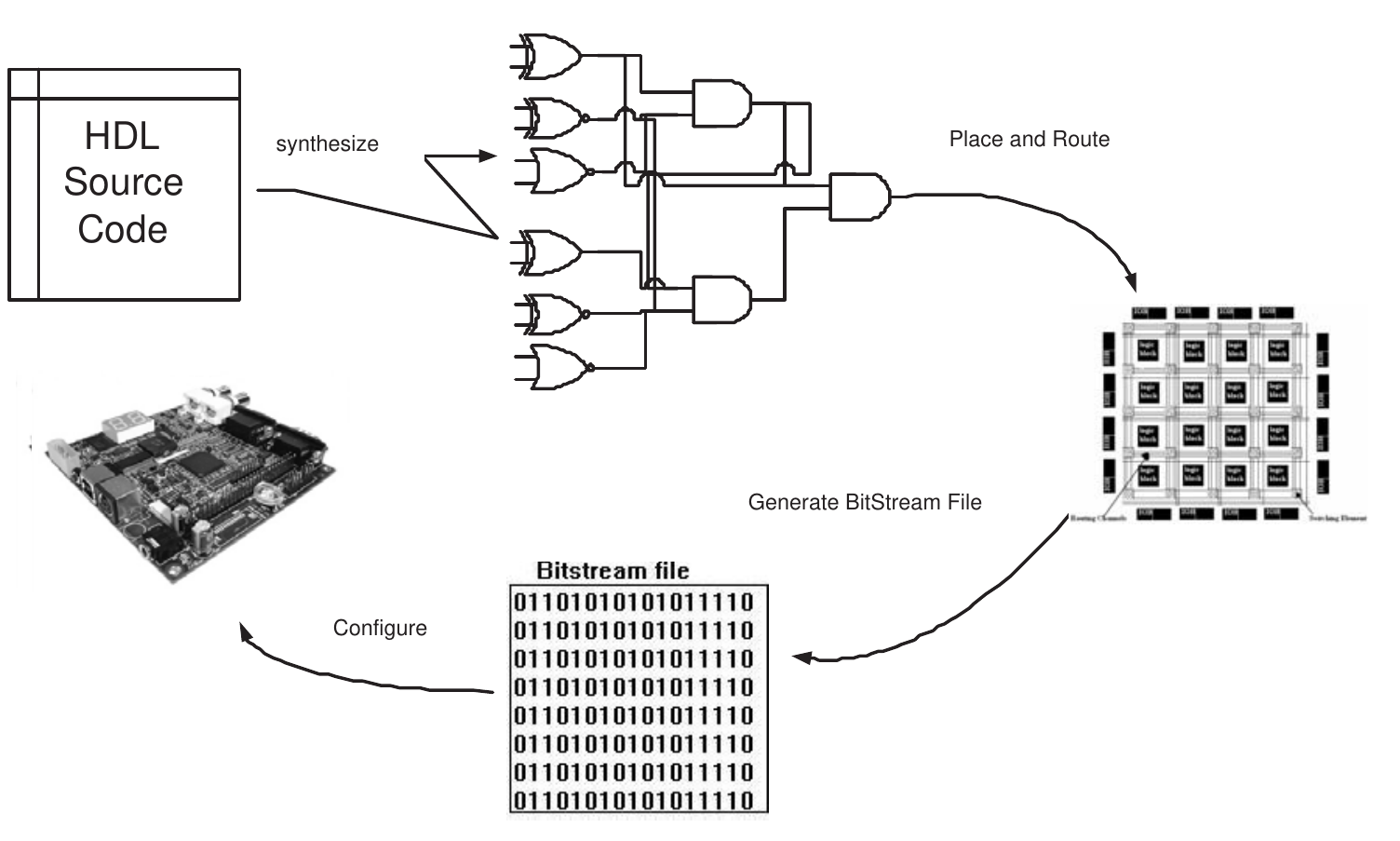}
\caption{\textit{FPGA} Design Flow \label{f3}}
\end{center}
\end{figure}
The above process consumes a remarkable amount of time; this is due
to the design that the user should provide using \textit{HDL}, most
probably \textit{VHDL} or \textit{Verilog}. The complexity of
designing in \textit{HDL}; which have been compared to the
equivalent of assembly language; is overcome by raising the
abstraction level of the design; this move is achieved by a number
of companies such as \textit{Celoxica}, \textit{Cadence} and
\textit{Synopsys}. These companies are offering higher level
languages with concurrency models to allow faster design cycles for
\textit{FPGAs} than using traditional \textit{HDLs}. Examples of
higher level languages are \textit{Handel-C}, \textit{SystemC}, and
\textit{Superlog} [31], [39].

\subsection{\textit{Handel-C} Language}

\textit{Handel-C} is a high level language for the implementation of
algorithms on hardware. It compiles program written in a
\textit{C}-like syntax with additional constructs for exploiting
parallelism [39]. The \textit{Handel-C} compiler comes packaged with
the \textit{Celoxica DK Design Suite} which also includes functions
and memory controller for accessing the external memory on the
\textit{FPGA}. A big advantage, compared to other \textit{C} to
\textit{FPGA} tools, is that \textit{Handel-C} targets hardware
directly, and provides a few hardware optimizing features [12]. In
contrast to other \textit{HDLs}, such as \textit{VHDL},
\textit{Handel-C} does not support gate-level optimization. As a
result, a \textit{Handel-C} design uses more resources on an
\textit{FPGA} than a \textit{VHDL} design and usually takes more
time to execute. In the following subsections, we describe
\textit{Handel-C} features' that we have used in our design [12],
[32].

\subsubsection{Types and Type Operator}

Almost all \textit{ANSI-C} types are supported in \textit{Handel-C}
with the exception of float and double. Yet, floating point
arithmetic can still be performed using the floating point library
provided by \textit{Celoxica}. Also, \textit{Handel-C} supports all
\textit{ANSI-C} storage class specifies and type qualifiers expect
volatile and register which have no meaning in hardware.
\textit{Handel-C} offers additional types for creating hardware
components such as memory, ports, buses and wires. \textit{Handel-C}
variables can only be initialized if they are global or if declared
as static or const.

\subsubsection{Values and Widths}

Unlike conventional \textit{C} types, \textit{Handel-C} types are
not limited to width since when targeting hardware, there is no need
to be tied to a certain width. Variables can be of different widths,
thus minimizing the hardware usage. For instant, if we have a
variable a that can hold a value between $1$ and $5$, then it is
enough to use $3$ bits only.

However, care should be taken when performing arithmetic or
comparisons on variables of different width. \textit{Handel-C}
offers the three operators: 1) concatenation '@' 2) take  '<-' and
3) drop '\\' for dealing with variables of different width. For
instant, if we want to add the variable a to another variable of
type long, we have to pad a to 32 bits by using the concatenation
operator @.

\subsubsection{\textit{par} Statement}

The notion of time in \textit{Handel-C} is fundamental. Each
assignment happens in exactly one clock cycle, everything else is
"free" [12].

An essential feature in \textit{Handel-C} is the 'par' construct
which executes instructions in parallel. Table \ref{table1} shows
the effect of using '\textit{par}'.

\begin{table}[h]
\caption{Effect of using '\textit{'par'} construct} \label{table1}
\begin{center}
\begin{tabular}{c c c| c c c c c}
\         && &&   par\\
\ a=1;    && &&    \{\\
\ b=1;    && &&    a=1\\
\ c=1;    && &&  b=1\\
\         && &&  c=1\\
\         && && \}\\
\         && &&\\
\ No. of clock cycles = 3 &&&& No. of clock cycles = 1
\end{tabular}
\end{center}
\end{table}

\subsubsection{\textit{Handel-C} Targets}

\begin{figure}[h]
\begin{center}
\includegraphics{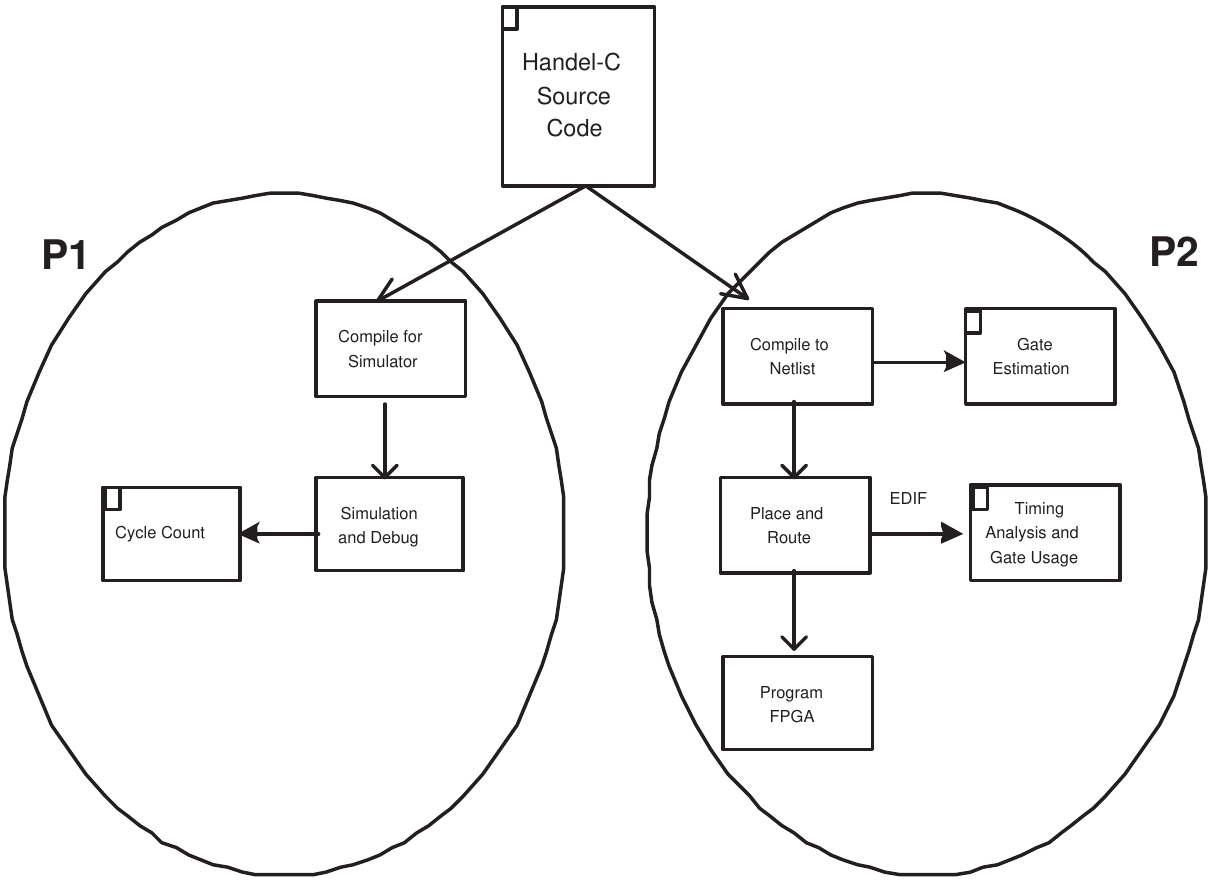}
\caption{\textit{Handel-C} targets \label{f4}}
\end{center}
\end{figure}

Handel-C supports two targets. The first is a simulator that allows
development and testing of code without the need to use hardware, P1
in Figure \ref{f4}. The second is the synthesis of a netlist for
input to place and route tools which are provided by the
\textit{FPGA's} vendors, P2 in Figure \ref{f4}.

The remaining of this section describes the phases involved in P2,
as it is clear from P1 that we can test and debug our design when
compiled for simulation.

The flow of the second target involves the following steps: Compile
to netlist: The input to this phase is the source code. A synthesis
engine, usually provided by the \textit{FPGA} vendor, translates the
original behavioral design into gates and flip flops. The resultant
file is called the netlist. Generally, the netlist is in the
Electronic Design Interchange Format \textit{(EDIF)} format. An
estimate of the logic utilization can be obtained from this phase.

Place and Route (\textit{PAR}): The input to this phase is the
\textit{EDIF} file generated from the previous phase; i.e. after
synthesis. All the gates and flip flops in the netlist are
physically placed and mapped to the \textit{FPGA} resources. The
\textit{FPGA} vendor tool should be used to place and route the
design. All design information regarding timing, chip area and
resources utilization are generated and controlled for optimization
at this phase.

Programming and configuring the \textit{FPGA}: After synthesis and
place and route, a binary file will be ready to be downloaded into
the \textit{FPGA} chip [15], [34].

\subsection{Field Programmable Gate Arrays}

An \textit{FPGA} is a programmable digital logic chip. It consists
of arrays of logic blocks with an interconnection network of wires.
Both the logic blocks and the interconnects can be programmed by the
designer so that the \textit{FPGA} can perform whatever logical
function is needed. Generally, the internal components of an
\textit{FPGA} can communicate with the outside world through the
Input/Output blocks (\textit{IOB}).

\subsubsection{\textit{FPGA} Structure}

Three important elements that characterize the architecture of any
\textit{FPGA} are: building/logic blocks, routing channels, and
switching elements. The great strength of \textit{FPGAs} is their
flexible architecture which enables the designer to program each of
these elements.

The building/logic blocks are the basic elements of an
\textit{FPGA}. Each of these blocks is configured to perform a logic
function. The interconnection between the logic blocks is provided
by the channels of wiring segments of varying lengths [13].  The
switching elements are used to determine the choice of active logic
modules and their interconnects. The designer can activate or
deactivate these elements to suite the requirement of the
application in hand [27].

There are several programming technologies for implementing the
programmable switches in an \textit{FPGA}; anti-fuse, \textit{SRAM}
memory and \textit{EPROM}-based technology. In the anti-fuse
technology, a strong electric current is used to create a connection
between the two-terminal anti-fuse device [33]. In the second
technology, \textit{SRAM} based configuration can be reprogrammed by
downloading different configuration bits into the \textit{SRAM}
memory cells [27]. The anti-fuse technology is faster and more
flexible than the \textit{SRAM}-based technology whose
programmability is volatile; i.e., the \textit{FPGA} needs to be
reconfigured whenever the power is turned off [28].

The architecture of an \textit{FPGA} can be classified according the
size and flexibility of the logic cell as well as to the structure
of the routing scheme [33]. The four basic architectures are:
Symmetrical array, Row-based, Fine Grain Cellular Architecture
(Sea-of-Gates), and Complex or Hierarchical \textit{PLD}.

\subsubsection{Reconfigurability of \textit{FPGA}}

\textit{FPGA} reconfiguration can be either static, semi static or
dynamic. The dynamic reconfiguration, also known as run-time
reconfiguration, is the most powerful form since a dynamically
reconfigurable \textit{FPGA} can be programmed / modified on-the-fly
while the system is operating [4]. Dynamically reconfigurable
\textit{FPGA} may be either partially reconfigured (local run time
reconfiguration) or programmed in a full reconfiguration (global run
time reconfiguration). In the first case, only a portion of the
\textit{FPGA} is \textit{"reconfigured while the other part is still
running. In the second case, all the system is reconfigured. An
external storage is needed to keep the intermediate results until
the configured functions run"} [36].

\section{Hardware Implementation of V-cycle \textit{MG}}

All available \textit{MG} solvers are realized as software running
on general purpose processors [30], [42].Available software packages
have been implemented in C, Fortran-77, Java and other languages,
where parallelized versions of these packages require
inter-processor communication standards such as Message Passing
Interface \textit{(MPI)} [23]. Each of these packages attempt to
achieve an efficient and a scalable version of the algorithm by
compromising between the accuracy of the solution and the speed of
realizing the solution.

The V-cycle \textit{MG, Jacobi, and SOR} algorithms have been
designed, implemented and simulated using \textit{Handel-C}. We have
targeted a \textit{Xilinx Virtex II Pro FPGA}, an \textit{Altera
Stratix FPGA}, and an \textit{RC10} board from \textit{Celoxica}.
The tools provided by the device's vendors were used to synthesize
and place and route the design [1], [12], [43].

Finding the solution to \textit{PDEs} using either of the
aforementioned techniques \textit{(MG, Jacobi, SOR)} requires
floating point arithmetic operations which are 1) far more complex,
and 2) consume more area than fixed point operations.  For this
reason, \textit{Handel-C} does not support floating point type. Yet,
floating point arithmetic can be performed using the Pipelined
Floating Point Library provided in the Platform Developer's Kit.

An unexpected crash in the \textit{Handel-C} simulator persists
whenever the numbers of floating point arithmetic operations exceed
four. The only possible way to avoid the simulator's failure was to
convert/Unpack the floating point numbers to integers and perform
integer arithmetic on the obtained unpacked numbers.Though it costs
more logic to be generated, the integer operations on the unpacked
floating point numbers have a minor effect on the total number of
the design's clock cycles.

\begin{figure}[h]
\begin{center}
\includegraphics{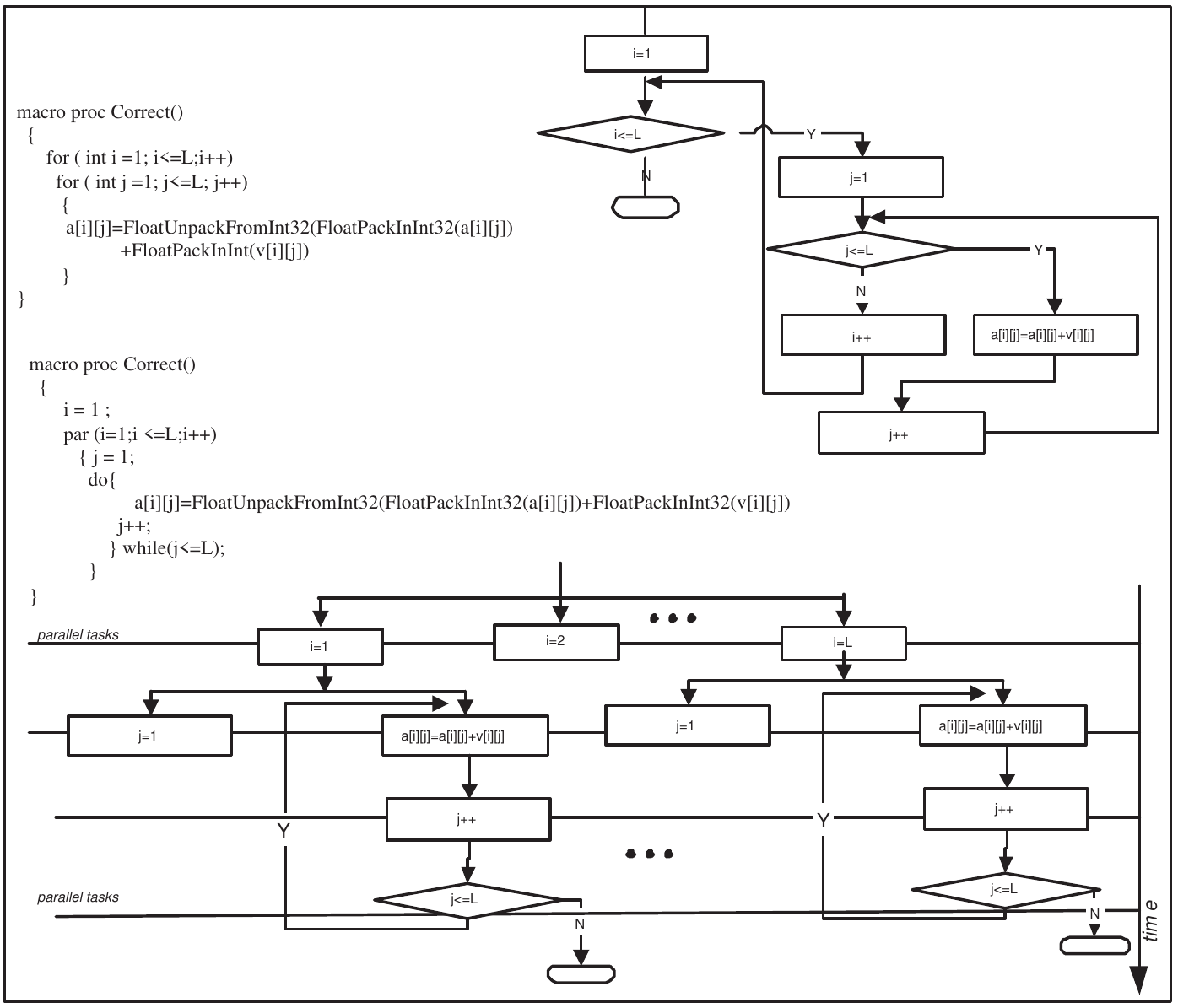}
\caption{\textit{MG} Correct operator, illustrating the effect of
using par construct: (6a), (6b), (6c) and (6d) shows sequential
code, flowcharts, parallel code and combined flowchart/concurrent
process model, respectively. The dots represent replicated instances
in d. Dashed lines show the parallel tasks\label{f6}}
\end{center}
\end{figure}

\begin{figure}[h]
\includegraphics{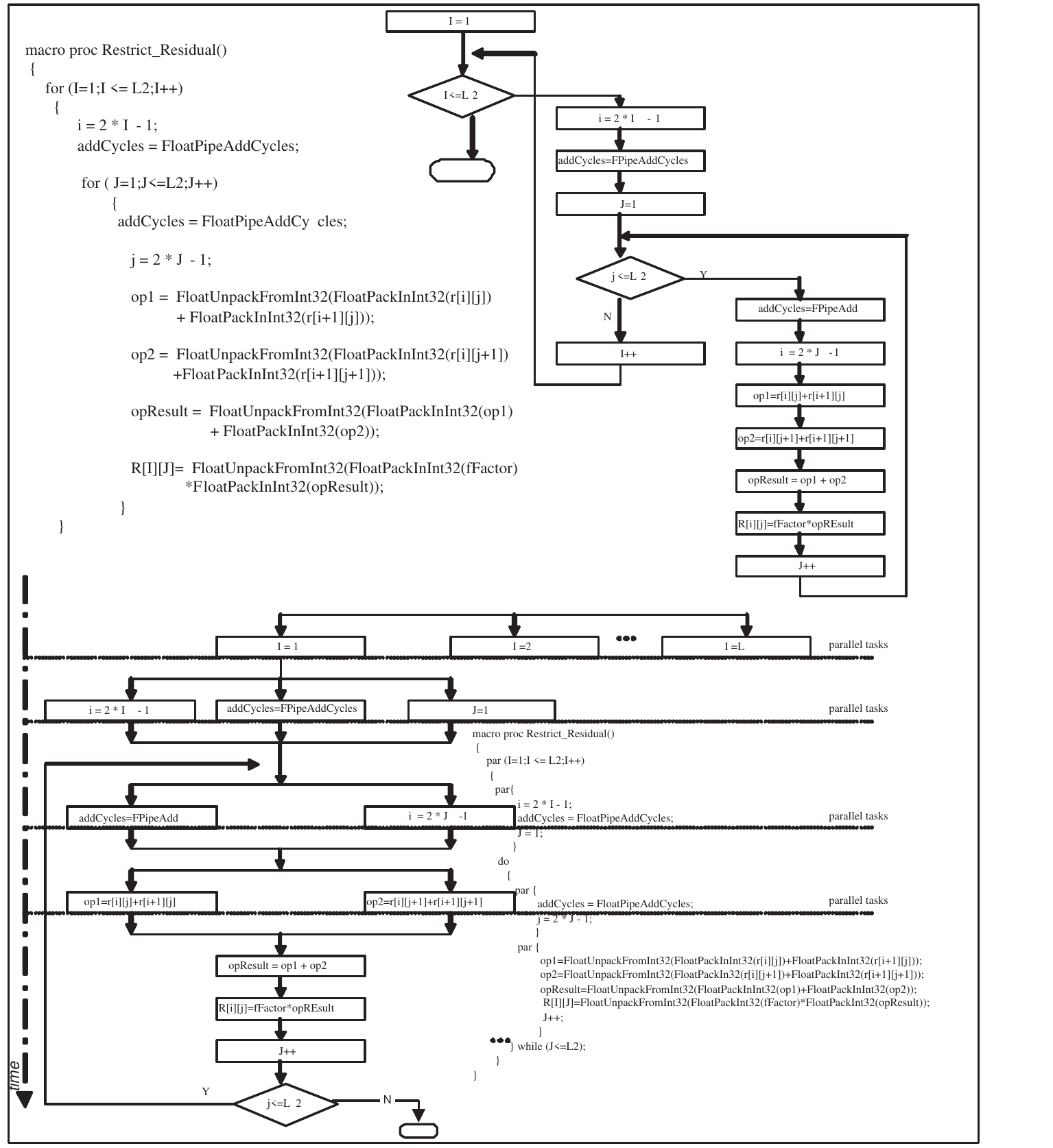}
\caption{\textit{MG} Restrict Residual operator, illustrating the
effect of using par construct: (7a), (7b), (7c) and (7d) shows
sequential code, flow charts, parallel code and combined flow
chart/concurrent process model, respectively. The dots represent
replicated instances in d). Dashed lines show the parallel
tasks.\label{f7}}
\end{figure}

\begin{figure}[h]
\begin{center}
\includegraphics{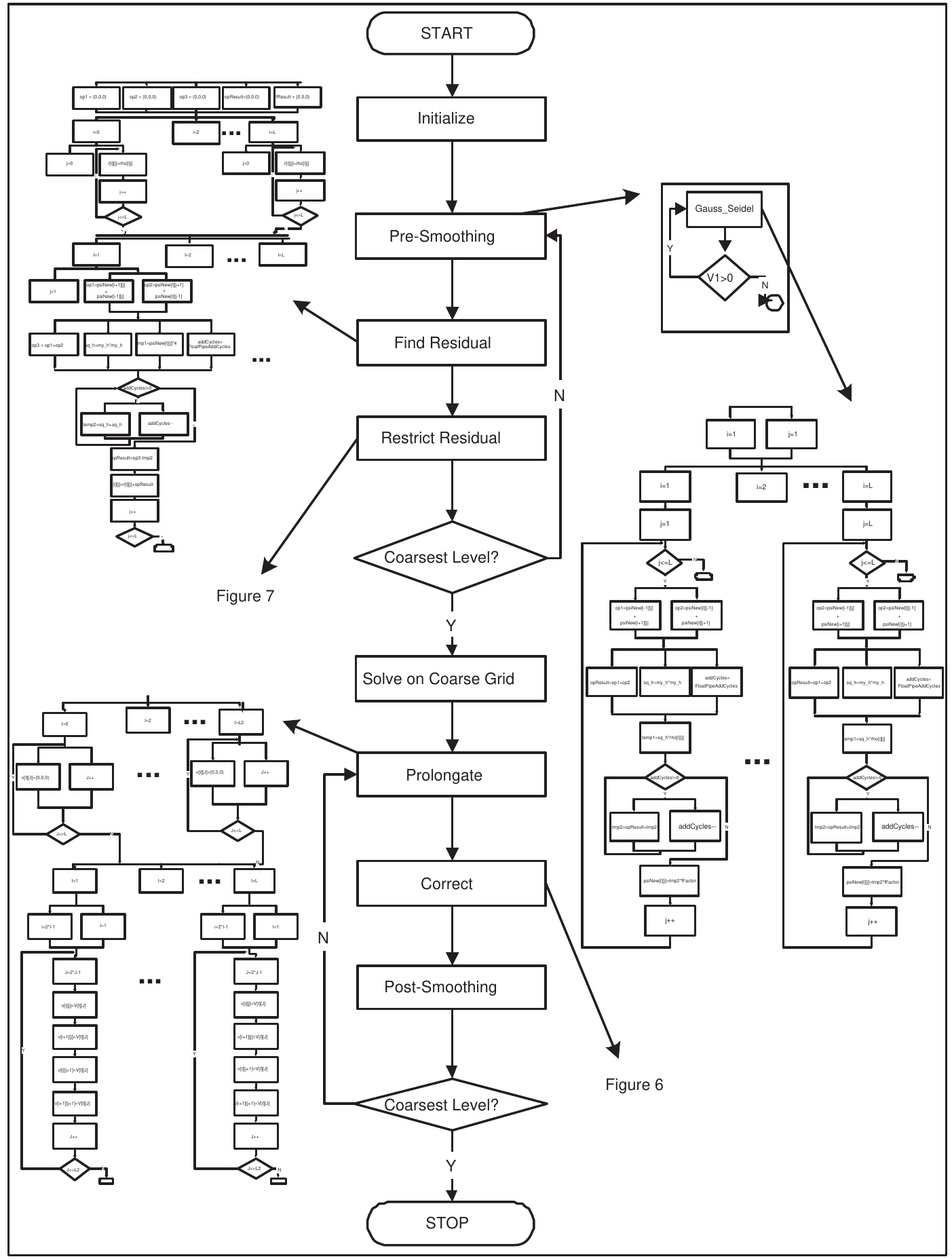}
\caption{V-cycle \textit{MG}, iterative version showing each
component parallelization. The dots in each of the component's
combined flowchart/concurrent process model represent replicated
instances.\label{f8}}
\end{center}
\end{figure}

The Multigrid method can be parallelized by parallelizing each of
its components; i.e., smoother, coarse grid solver, restriction and
prolongation. Each of these components is parallelized by using the
\textit{Handel-C} construct '\textit{par}'. This is used whenever it
was possible to execute more than one instruction in parallel
without affecting the logic of the source code. Figures \ref{f6} and
\ref{f7} show the two \textit{MG} operators 'Restrict Residual' and
'Correct'. A snapshot of the parallel version of the 'Smoother',
'Find Residual' and 'Prolongate' components is shown in Figure
\ref{f8}. Their implementation style is very similar to that of
'Restrict Residual' and 'Correct' operators.

Both \textit{Jacobi} and \textit{SOR} methods have been parallelized
in the same way, i.e., using the par construct whenever possible.
The results obtained show: a)  The robustness of \textit{MG}
algorithm over \textit{Jacobi} and \textit{SOR} algorithm in both
hardware and software implementations. b)  A substantial improvement
in the \textit{MG}, \textit{Jabobi}, and \textit{SOR} performance
when compared to the traditional way of executing instructions on a
\textit{GPP}.

\section {Experimental Results}
The \textit{Handel-C} simulators along with the \textit{FPGA}
vendor's tools were used to obtain the results. We draw a comparison
of the execution time between our results and a software version
written in \textit{C++}. The obtained results are based on the
following criteria:

\begin{itemize}
\item \textbf{Speed of convergence}: the time it takes the method of choice to
find the solution to the \textit{PDE} in hand. In another word, it
is the time needed to execute \textit{MG}, \textit{Jacobi} or
\textit{SOR} algorithm. In hardware implementation, the speed of
convergence is measured using the clock cycles of the design divided
by the frequency at which the design operates at. The first
parameter is found using the simulator while the second is found
using the timing analysis report which is generated using the
\textit{FPGA} vendor's tool.

\item \textbf{Accuracy of the solution}: The convergence of each algorithm
is greatly dependent on the accuracy of the solution. The increase
in the accuracy results in the increase of the computation as well
as the increase in the logic utilization.

\item \textbf{chip-area}: this performance criterion measures the number of
occupied slices on the \textit{FPGA} on which the design is
implemented. The number of occupied slices is generated using the
\textit{FPGA} vendor's place and route tool.

\end {itemize}

We compare the timing performance between our hardware
implementations of Multigrid, Jacobi, \textit{SOR} and a
\textit{C++} software version of the same algorithms on GPPs.

The following selections were used for all Multigrid performance
tests: \begin{itemize}

\item Restriction: Full Weighting,
\item Interpolation: Bilinear,
\item Number of smoothing steps:
\item Smoother used: Gauss-Seidel
\item Accuracy: 0.001 for all \textit{Handel-C} test cases and \textit{C++} test cases
up to problem size 64x64.

\end {itemize}

As for \textit{SOR} performance tests, the over-relaxation
parameters, omega, is set to be 1.5.

The V-cycle \textit{MG} execution time, for different problem sizes,
along with the maximum frequency at which each design operates at
are shown in Table \ref{table1}. The execution time is calculated
using: No. of clock cycles/Max. Frequency.

\begin{table}[h]
\caption{Execution Time and Max frequency for different problem
sizes} \label{table1}
\begin{center}
\begin{tabular}{|c| c| c|}
\hline Mesh Size & Execution Time& Fmax\\
\hline
\hline 8x8 &      0.000063    &       159.74\\
\hline 16x16 &    0.00026    &      153.52\\
\hline 32x32 &       0.00118 &     136.15\\
\hline 64x64 &      0.00555  &     115.97\\
\hline 128x128 &      0.031  &    83.91 \\
\hline 256x256 &       0.188 &  54.60  \\
\hline 512x512 &    1.308    &      31.45\\
\hline 1024x1024 &    9.3    &    17.60 \\
\hline 2048x2048 &     70.97   &     9.28 \\
\hline
\end{tabular}
\end{center}
\end{table}

\begin{figure}[h]
\begin{center}
\includegraphics{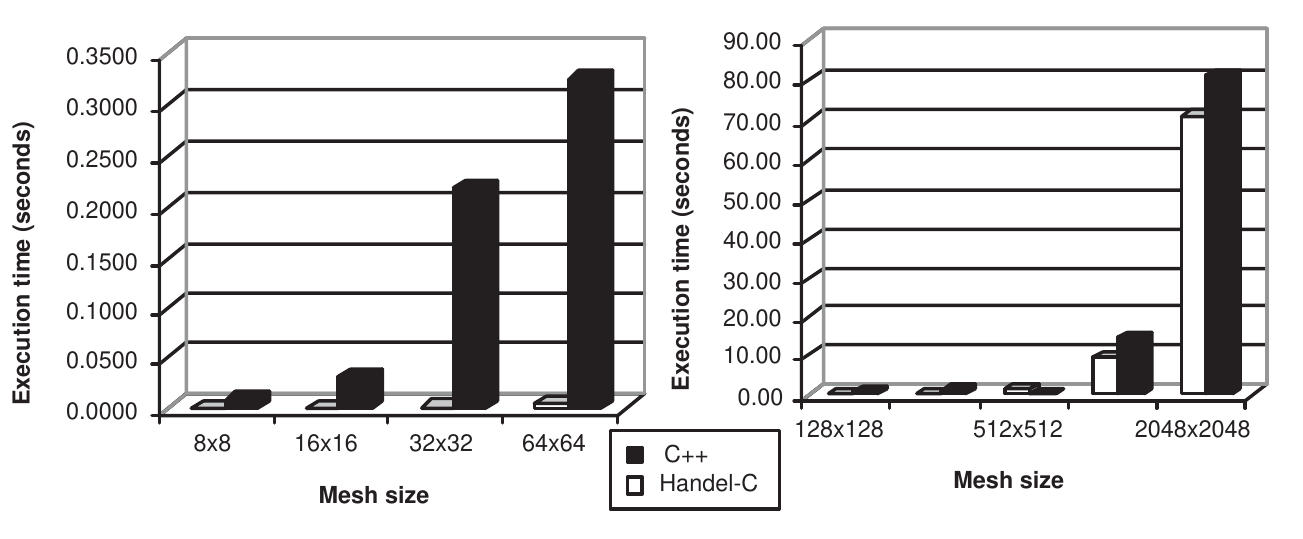}
\caption{\textit{MG} execution time results in both versions,
\textit{Handel-C} and \textit{C++}\label{f9}}
\end{center}
\end{figure}

Figure \ref{f9} shows the results of comparing the execution time
when running a \textit{C++} version of the V-cycle Multigrid
algorithm and our proposed \textit{Handel-C} version. The
superiority of the hardware implementation over the software
implementation is clear in both figures. However, for a problem size
greater than 64x64, it becomes difficult to measure the execution
time of the software (\textit{C++}) version with the same accuracy
of 0.001. At that time, our concern was to force the \textit{C++}
version of \textit{MG} to converge at any price. This was only
possible by sacrificing with the accuracy of the solution; where we
had to gradually increase this factor until we reached an accuracy
of 2.0 for a problem size of 2048x2048, in contrast to an accuracy
of 0.001 for a problem size of 8x8. On the other hand,
\textit{Handel-C} results were independent from the accuracy of the
solution. The accuracy was constant all the way from a problem size
of 8x8 to 2048x2048. Obviously, this explains the degeneration of
the speedup indicated in (b).

In Table \ref{table2} we draw a comparison between the accuracy of
the solution for each of the \textit{C++} and \textit{Handel-C} test
cases. The speedup of the design is calculated as the ratio of
Execution Time (\textit{C++}) / Execution Time (\textit{Handel-C}).

\begin{table}[h]
\caption{Required accuracy of the solution for \textit{C++} and
\textit{Handel-C} test cases, and the designs speedup}\label{table2}
\begin{center}
\begin{tabular}{|c|c|c|c|c|c|c|}
\hline Mesh Size &\multicolumn{2}{c|}{Accuracy}&\multicolumn{3}{|c|}{Speedup}\\
\cline{2-6}
 &\textit{C++}&\textit{Handel-C}& \textit{MG} & \textit{SOR} &\textit{Jacobi}\\
\hline\hline
8x8&0.001&0.001&142.86&1.758&223.81\\
16x16&0.001&0.001&185.59&188&56.21\\
32x32&0.001&0.001&119.23&6.706&5.68\\
64x64&0.001&0.001&58.56&5.69&2.89\\
128x128&0.001&0.001&20.77&1.514&1.41\\
256x256&1&0.001&5.25&1.43&2.29\\
512x512&1.1&0.001&2.92&3.03&2.39\\
1024x1024&1.3&0.001&1.58&2.58&0.75\\
2048x2048&2&0.001&1.14&3.37&1.39\\
\hline
\end{tabular}
\end{center}
\end{table}

\begin{figure}[h]
\begin{center}
\includegraphics{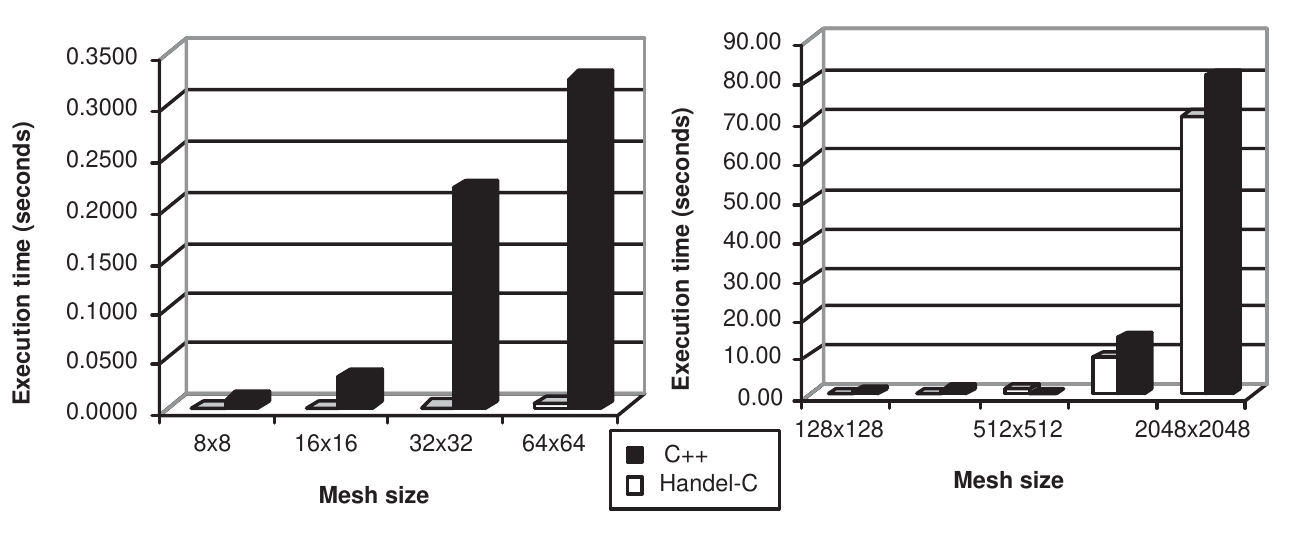}
\caption{Jacobi execution time results in both versions,
\textit{Handel-C} and \textit{C++}\label{f10}}
\end{center}
\end{figure}

\begin{figure}[h]
\begin{center}
\includegraphics{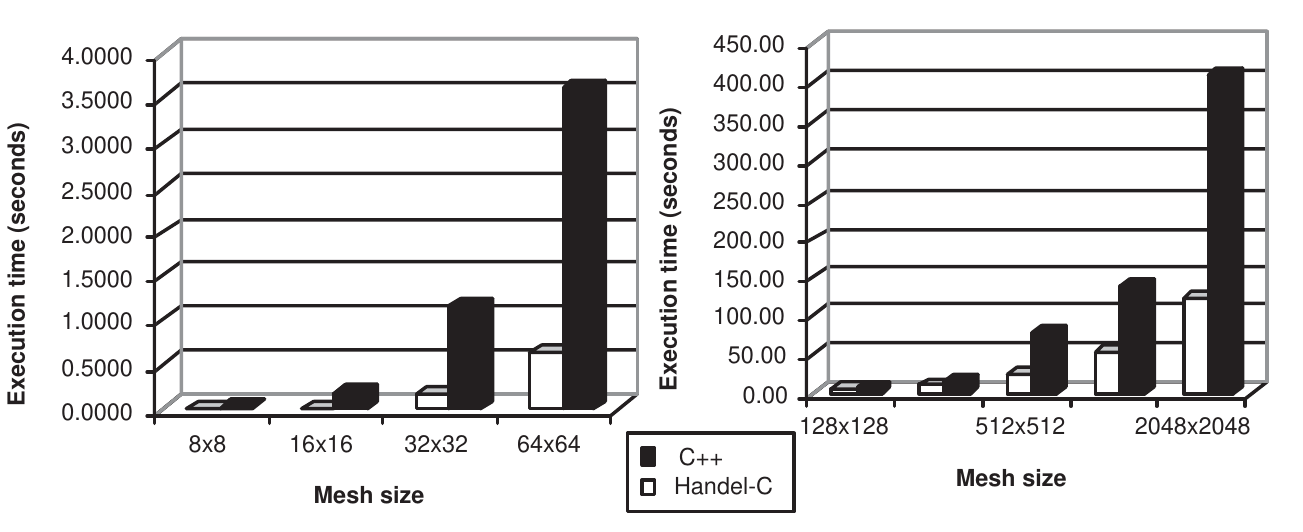}
\caption{\textit{SOR} execution time results in both versions,
\textit{Handel-C} and \textit{C++}\label{f11}}
\end{center}
\end{figure}

The superiority of the hardware implementation over the software
implementation for \textit{Jacobi} and \textit{SOR} is shown in
Figures \ref{f10} and \ref{f11}. This observation demonstrates the
ability of realizing an accelerated version of the algorithm when
implemented on hardware.

\begin{figure}[h]
\begin{center}
\includegraphics{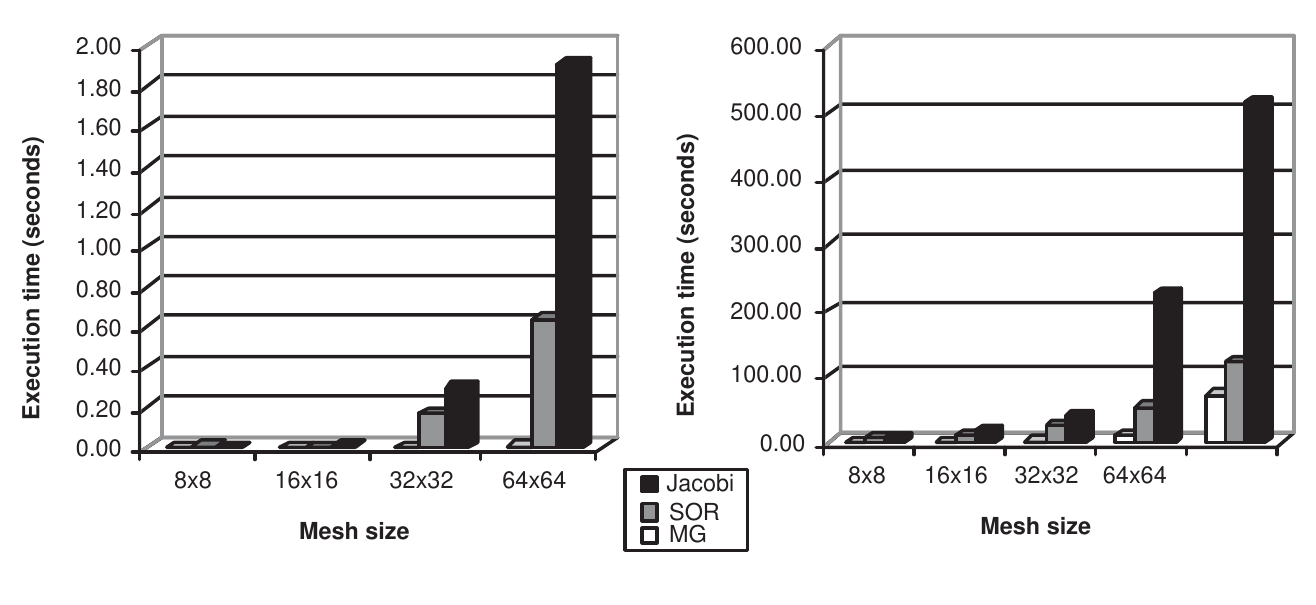}
\caption{Robustness of \textit{MG} over \textit{Jacobi} and
\textit{SOR}\label{f12}}
\end{center}
\end{figure}

Tables \ref{table3}, \ref{table4}, \ref{table5} show, respectively,
the \textit{Virtex II Pro.}, \textit{Spartan3L} and \textit{Altera
Stratix FPGA} synthesis results for different problem sizes in
\textit{MG,SOR}, and \textit{Jacobi}. When targeting \textit{Xilinx
Virtex II Pro FPGA}, the largest possible problem size that we could
achieve was 2048x2048,where 99\% of the slices were utilized.
Meanwhile, the largest possible problem size was 512x512 when
targeting Spartan3L \textit{FPGA}.

\begin{table}[h]
\caption{\textit{Xilinx Virtex II Pro} Synthesis Results using
\textit{Xilinx ISE}} \label{table3}
\begin{center}
\begin{tabular}{|c|c|c|c|c|c|c|c|c|}
\hline Mesh Size &\multicolumn{3}{|c|}{Number of Occupied
Slices}&\multicolumn{3}{|c|}{Total equivalent gate
count}\\\hline\hline
&\textit{MG}&\textit{SOR}& \textit{Jacobi} & \textit{MG}& \textit{SOR} &\textit{Jacobi}\\
\hline 8x8&264&128&146&5990&2918&3229\\
\hline 16x16&295&136&159&6497&3033&3397\\
\hline 32x32&415&219&299&9321&4807&5090\\
\hline 64x64&536&265&380&12376&5978&7849\\
\hline 128x128&789&315&499&18107&7125&11864\\
\hline 256x256&1247&610&839&29244&14538&17864\\
\hline 512x512&2125&1098&1286&51115&23012&23649\\
\hline 1024x1024&3875&1601&1890&94484&31848&31327\\
\hline 2048x2048&4926&2289&3198&180879&53476&35839\\
\hline \hline
\end{tabular}
\end{center}
\end{table}

\begin{table}[h]
\caption{\textit{Spartan3L} Synthesis Results using \textit{Xilinx
ISE}} \label{table4}
\begin{center}
\begin{tabular}{|c|c|c|c|c|c|c|c|c|}
\hline Mesh Size &\multicolumn{3}{|c|}{Number of Occupied
Slices}&\multicolumn{3}{|c|}{Total equivalent gate
count}\\\hline\hline
&\textit{MG}&\textit{SOR}& \textit{Jacobi} & \textit{MG}& \textit{SOR} &\textit{Jacobi}\\
\hline 8x8&687&302&416&355687&279010&356109\\
\hline 16x16&717&499&599&356163&281001&357631\\
\hline 32x32&769&589&7326&357224&282997&359989\\
\hline 64x64&832&745&9010&358921&284000&342768\\
\hline 128x128&1049&877&1198&361956&285872&389999 \\
\hline 256x256&1507&1201&1665&367673&297134&397987  \\
\hline 512x512&3187&2010&2810&375293&299858&498030\\
\hline
\end{tabular}
\end{center}
\end{table}

\begin{table}[h]
\caption{\textit{Altera Stratix} Synthesis Results using
\textit{Quartus II}} \label{table5}
\begin{center}
\begin{tabular}{|c|c|c|c|c|c|c|c|c|c|}
\hline Mesh Size &\multicolumn{3}{|c|}{Total Logic
Elements}&\multicolumn{3}{|c|}{\scriptsize LE Usage by No. of LUT
Inputs}&\multicolumn{3}{|c|}{Total Registers}\\\hline
\hline &\textit{MG}&\textit{SOR}& \textit{Jacobi} & \textit{MG}& \textit{SOR} &\textit{Jacobi}& \textit{MG}& \textit{SOR} &\textit{Jacobi}\\
\cline{2-4} \hline 8x8&725&519&610&402&250&354&228&120&189\\
\hline 16x16     &   818&601&709&554&310&401&265&155&232\\
\hline 32x32     &   925&810&880&625&501 & 556&301&199&300\\
\hline 64x64     &   1068&999&1001&709&637 & 681&360&280&385\\
\hline 128x128   &   1307&1274&1286&841&720 & 801&467&347&390\\
\hline 256x256   &   1739&1510&1590&1070&890 & 950&670&498&476\\
\hline 512x512   &   2653&2286&2589&1357&1087 & 1101&816&501&560\\
\hline 1024x1024 &   3491&2901&3342&1809&1450 & 1499&1002&569&689\\
\hline 2048x2048 &   4501&3286&3927&2201&1798 & 1941&482&640&819\\
\hline
\end{tabular}
\end{center}
\end{table}

\section{Conclusions and Future Work}
In this paper, we have presented a hardware implementation of the
V-cycle Multigrid method for solving the Poisson equation in two
dimensions. \textit{Handel-C} hardware compiler is used to code and
implement our designs (\textit{MG}, \textit{Jacobi}, and
\textit{SOR}) and map them onto high-performance \textit{FPGAs},
such as, \textit{Virtex II Pro}, \textit{Altera Stratix}, and
\textit{Spartan3L} which is embedded in the \textit{RC10 FPGA}-based
system from \textit{Celoxica}. The implementation performance is
analyzed using the \textit{FPGAs} vendors' proprietary software.
Moreover, we compare our implementation results with available
software version results running on General Purpose Processors and
written in C++. The obtained results have demonstrated that 1)
\textit{MG} algorithm outperforms the \textit{Jacobi} and the
\textit{SOR} algorithms, on both hardware and software and 2)
\textit{MG} on hardware outperforms \textit{MG} on \textit{GPP},
where a speedup of 142.86 was achieved for a problem size of 8x8,
whereas a speedup of 1.14 was achieved for 2048x2048. This
degeneration of the speedup is due to the increase of the value of
the required accuracy of the solution. Possible future directions
include realizing a pipelined version of the algorithm, moving to a
lower-level \textit{HDL} such as \textit{VHDL}, mapping the
algorithm into a coarse grain reconfigurable systems (e.g.,
\textit{MorphoSys}) [28], and benefiting from the advantages of
formal modeling [29]. We can also extend the benefit of \textit{MG}
by implementing the W-cycle algorithm and the Algebraic \textit{MG}.

\end{document}